\pdfoutput=1

\documentclass[11pt]{article}

\usepackage{EMNLP2022}

\usepackage{times}
\usepackage{latexsym}

\usepackage[T1]{fontenc}

\usepackage[utf8]{inputenc}

\usepackage{microtype}

\usepackage{inconsolata}

\usepackage{bm}
\usepackage{array}
\usepackage{xspace}
\usepackage{amssymb}
\usepackage{amsmath}
\usepackage{enumitem}
\usepackage{booktabs}
\usepackage{graphicx}
\usepackage{multirow}
\usepackage{multicol}
\usepackage{makecell}
\usepackage{subfigure}
\usepackage[normalem]{ulem}
\usepackage[ruled]{algorithm2e}

\usepackage{CJKutf8}
\usepackage[russian,english]{babel}

\setlist[itemize]{leftmargin=5mm, itemsep=0mm}

\newcommand{\ignore}[1]{}
\newcommand{\name}{QuiCK\xspace}
\newcommand{\ie}{\emph{i.e.,}\xspace}
\newcommand{\eg}{\emph{e.g.,}\xspace}

\newcommand{\aka}{\emph{a.k.a.,}\xspace}

\newcommand{\paratitle}[1]{\vspace{1ex}\noindent{\bf #1}}

\usepackage{xcolor}
\usepackage{color, soul}

\title{Empowering Dual-Encoder with Query Generator for \\ Cross-Lingual Dense Retrieval}

\author{
  Houxing Ren$^1$\thanks{~~Work done during internship at Microsoft STCA.} \quad Linjun Shou$^2$ \quad Ning Wu$^2$ \quad Ming Gong$^2$ \quad Daxin Jiang$^2$\thanks{~~Corresponding author.} \\
  $^1$School of Computer Science and Engineering, Beihang University \\
  $^2$Microsoft STC Asia \\
  renhouxing@buaa.edu.cn~~~~\{lisho,wuning,migon,djiang\}@microsoft.com
}

\begin{document}
\maketitle

\begin{abstract}
In monolingual dense retrieval, lots of works focus on how to distill knowledge from cross-encoder re-ranker to dual-encoder retriever and these methods achieve better performance due to the effectiveness of cross-encoder re-ranker. However, we find that the performance of the cross-encoder re-ranker is heavily influenced by the number of training samples and the quality of negative samples, which is hard to obtain in the cross-lingual setting. In this paper, we propose to use a query generator as the teacher in the cross-lingual setting, which is less dependent on enough training samples and high-quality negative samples. In addition to traditional knowledge distillation, we further propose a novel enhancement method, which uses the query generator to help the dual-encoder align queries from different languages, but does not need any additional parallel sentences. The experimental results show that our method outperforms the state-of-the-art methods on two benchmark datasets.
\end{abstract}

\section{Introduction} \label{sec:intro}

Information Retrieval~(IR) aims to retrieve pieces of evidence for a given query. 
Traditional methods mainly use sparse retrieval systems such as BM25~\cite{robertson2009probabilistic}, which depend on keyword matching between queries and passages. 
With the development of large-scale pre-trained language models~(PLMs)~\cite{vas2017attention,devlin2018bert} such as BERT, dense retrieval methods~\cite{lee2019latent,karpukhin2020dense} show quite effective performance. 
These methods usually employed a dual-encoder architecture to encode both queries and passages into dense embeddings and then perform approximate nearest neighbor searching~\cite{johnson2019billion}.

Recently, leveraging a cross-encoder re-ranker as the teacher model to distill knowledge to a dual-encoder has shown quite effective to boost the dual-encoder performance. Specifically, these methods first train a warm-up dual-encoder and a warm-up cross-encoder. Then, they perform knowledge distillation from the cross-encoder to the dual-encoder by KL-Divergence or specially designed methods. For example, RocketQAv2~\cite{qu2021rocketqa} proposed dynamic distillation, and AR2~\cite{zhang2021adver} proposed adversarial training.

\begin{figure}[t]
    \centering
    \subfigure[BM25.]{
        \includegraphics[width=0.46\columnwidth]{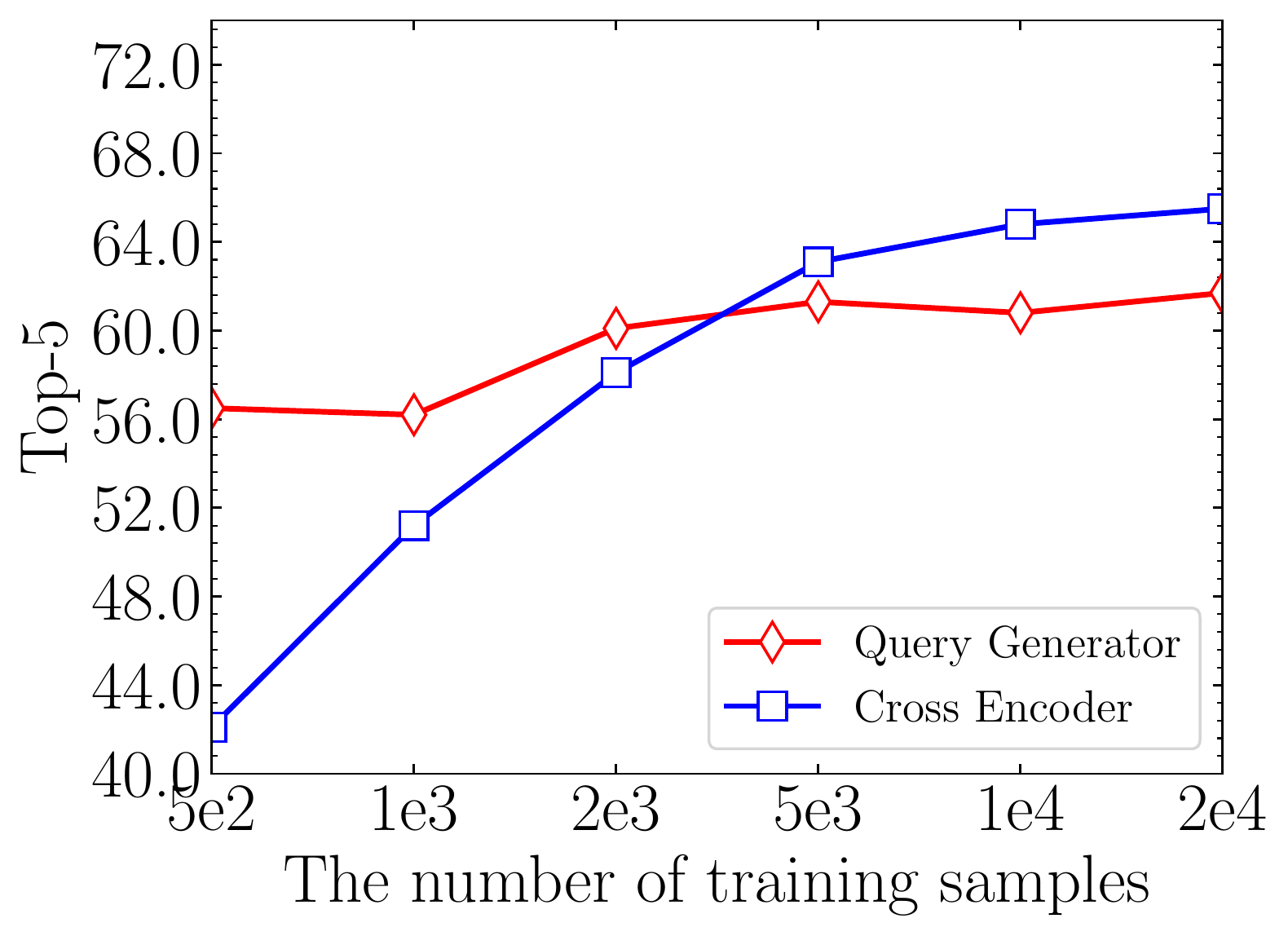}
    }
    \subfigure[DPR.]{
        \includegraphics[width=0.46\columnwidth]{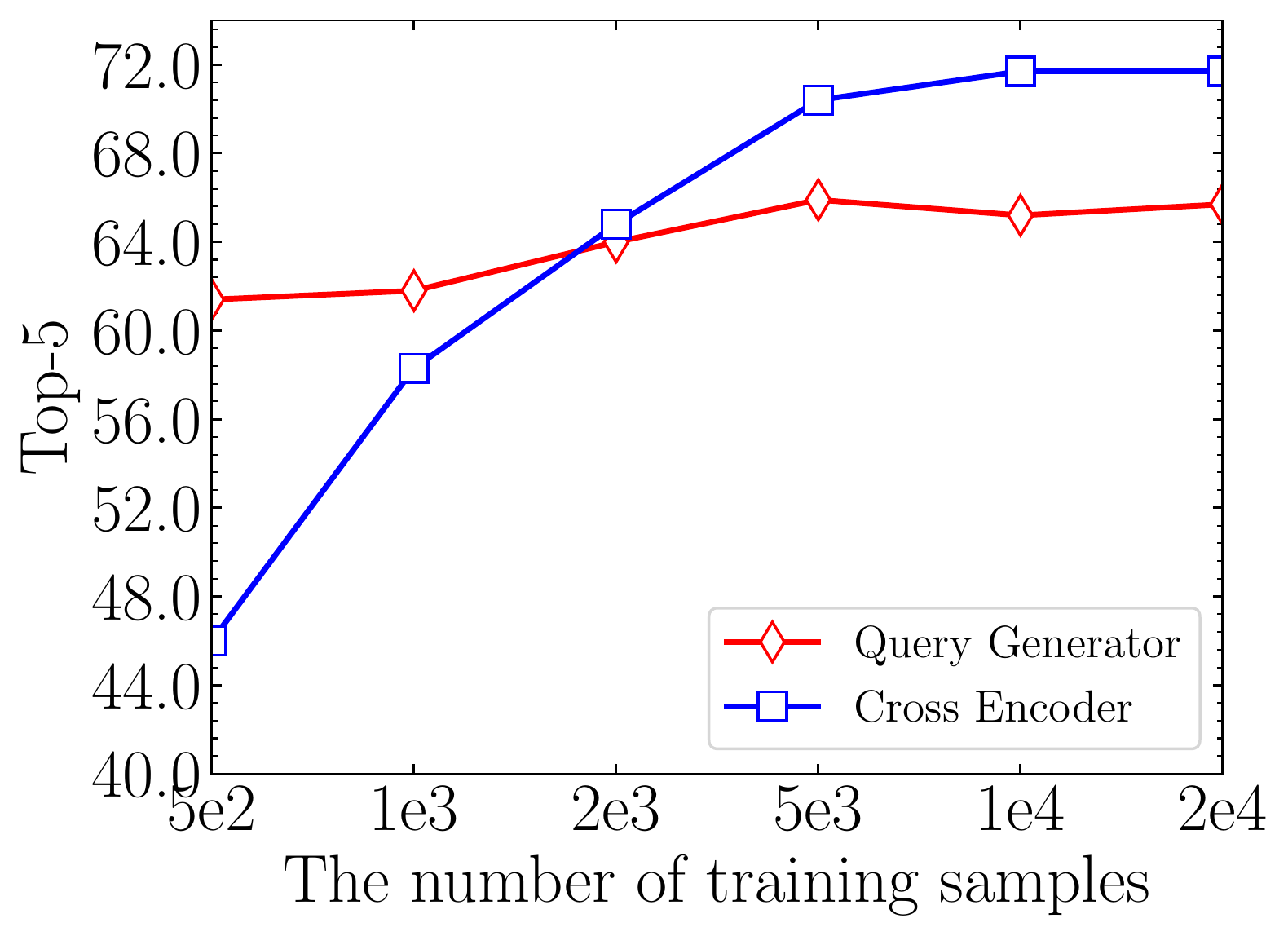}
    }
    \caption{The performance of cross-encoder and query generator when varying the number of training samples and retrievers. We use BM25 and DPR as retrievers, respectively. For the cross-encoder~(BERT-Large), we use retrieved top-100 passages which do not contain the answer as negative and contrastive loss for training. For the query generator~(T5-Base), we firstly train it with the query generation task and then fine-tune the model with the same setting as BERT-Large. The reported performance is the top-5 score of re-ranked top 500 passages on the NQ test set.}
    \label{fig:re-rank}
\end{figure}

However, there are two major problems when scaling the method to the cross-lingual dense retrieval setting.
Firstly, the cross-encoder typically requires large amounts of training data and high-quality negative samples  due to the gap between pre-training (token-level task) and fine-tuning (sentence-level task), which are usually not satisfied in the cross-lingual setting~\cite{asai2020xor}.
Due to expensive labeling and lack of annotators in global languages, especially low-resource languages, the training data in cross-lingual are quite limited. Then with the limited training data, the dual-encoder is not good enough to provide high-quality negative samples to facilitate the cross-encoder. 
Secondly, the cross-lingual gaps between different languages have a detrimental effect on the performance of cross-lingual models. Although some cross-lingual pre-training methods such as InfoXLM~\cite{chi2020infoxlm} and LaBSE~\cite{feng2022language} have put lots of effort into this aspect by leveraging parallel corpus for better alignment between different languages, these parallel data are usually expensive to obtain and the language alignment could be damaged in the fine-tuning stage if without any constraint.

To solve these problems, we propose to employ a query generator in the cross-lingual setting, which uses the likelihood of a query against a passage to measure the relevance. 
On the one hand, the query generator can utilize pre-training knowledge with small training data in fine-tuning stage, because both of its pre-training and fine-tuning have a consistent generative objective. 
On the other hand, the query generation task is defined over all tokens from the query rather than just the \emph{[CLS] token} in the cross-encoder, which has been demonstrated to be a more efficient training paradigm~\cite{clark2020electra}. As shown in Figure~\ref{fig:re-rank}, with the number of training samples dropping, the performance of BERT-Large drops more sharply than T5-Base. 
Besides, the query generator is less sensitive to high-quality negative samples. As we can see, using BM25 as the retriever to mine negative samples for re-ranker training, the gap between cross-encoder and query generator is smaller than the gap using DPR as the retriever. Finally, the query generator can provide more training data by generation, which is precious in the cross-lingual setting. To sum up, the query generator is more effective than the cross-encoder in the cross-lingual setting.

Based on these findings, we propose a novel method, namely \name, which stands \uline{Qu}ery generator \uline{i}mproved dual-encoder by \uline{C}ross-lingual \uline{K}nowledge distillation. Firstly, at the passage level, we employ a query generator as the teacher to distill the relevant score between a query and a passage into the dual-encoder. Secondly, at the language level, we use the query generator to generate synonymous queries in other languages for each training sample and align their retrieved results by KL-Divergence. Considering the noise in the generated queries, we further propose a scheduled sampling method to achieve better performance.

The contributions of this paper are as follows:
\begin{itemize}
    \item We propose a cross-lingual query generator as a teacher model to empower the cross-lingual dense retrieval model and a novel iterative training approach is leveraged for the joint optimizations of these two models.
    \item On top of the cross-lingual query generator, a novel cost-effective alignment method is further designed to boost the dense retrieval performance in low-resource languages, which does not require any additional expensive parallel corpus.
    \item Extensive experiments on two public cross-lingual retrieval datasets demonstrate the effectiveness of the proposed method.
\end{itemize}

\section{Related Work}

\paratitle{Retrieval.} 
Retrieval aims to search relevant passages from a large corpus for a given query. Traditionally, researchers use bag-of-words~(BOW) based methods such as TF-IDF and BM25~\cite{robertson2009probabilistic}. These methods use a sparse vector to represent the text, so we call them sparse retrievers. Recently, some studies use neural networks to improve the sparse retriever such as DocTQuery~\cite{nogueira2019doc2query} and DeepCT~\cite{dai2019context}. 

In contrast to sparse retrievers, dense retrievers usually employ a dual-encoder to encode both queries and passages into dense vectors whose lengths are much less than sparse vectors. These methods mainly focus on two aspects: pre-training~\cite{lee2019latent,guu2020realm,lu2021less,gao2021condenser,gao2021unsupervised,zhou2022hyperlink} and fine-tuning methods, including negative sampling~\cite{karpukhin2020dense,luan2020sparse,xiong2020approximate,zhan2021optimizing} and multi-view representations~\cite{khattab2020colbert,humeau2020poly,tang2021improving,zhang2022multi}. 
Another fine-tuning method is jointly training the dual-encoder with a cross-encoder. For example, RDR~\cite{yang2020retriever} and FID-KD~\cite{izacard2020distilling} distill knowledge from a reader to the dual-encoder;
RocketQA~\cite{qu2021rocketqa}, PAIR~\cite{ren2021pair}, RocketQAv2~\cite{ren2021rocketqav2}, and AR2~\cite{zhang2021adver} jointly train the dual-encoder with a cross-encoder to achieve better performance. 

\begin{figure*}[t]
    \centering
    \subfigure[Dual-Encoder.]{
        \includegraphics[width=0.6\columnwidth, page=1]{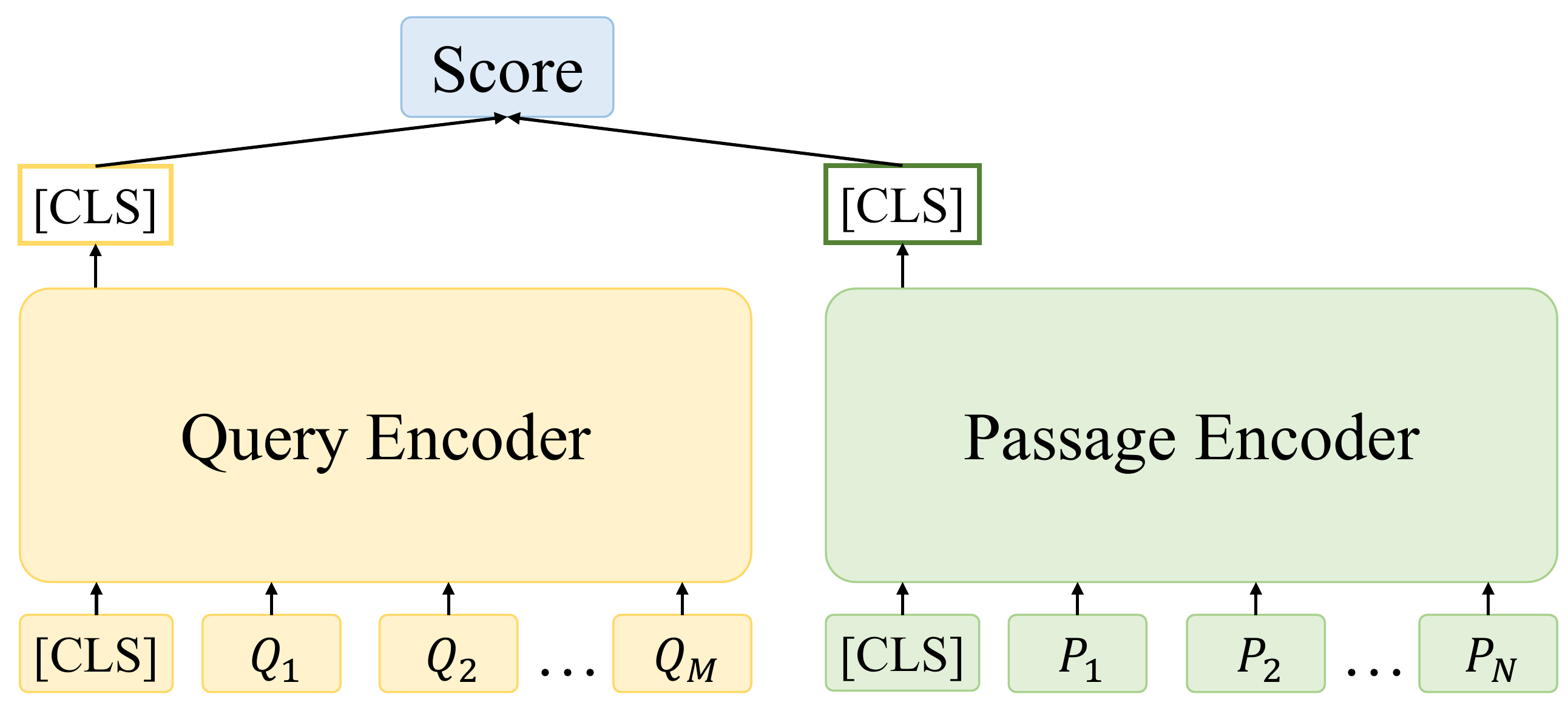}
    }
    \hspace{0.3cm}
    \subfigure[Cross-Encoder.]{
        \includegraphics[width=0.6\columnwidth, page=2]{figures/encoder.pdf}
    }
    \hspace{0.3cm}
    \subfigure[Query Generator.]{
        \includegraphics[width=0.6\columnwidth, page=3]{figures/encoder.pdf}
    }
    \caption{Overview of different model architectures designed for retrieval or re-ranking.}
    \label{fig:encoder}
\end{figure*}

Recently, with the development of cross-lingual pre-trained models~\cite{conneau2019unsupervised}, researchers pay more attention to cross-lingual dense retrieval~\cite{asai2020xor,longpre2020mkqa}. For example, CORA~\cite{asai2021one} leverages a generator to help mine retrieval training data, Sentri~\cite{soro2022ask} proposes a single encoder and self-training, and DR.DECR~\cite{li2021learning} uses parallel queries and sentences to perform cross-lingual knowledge distillation. 

\paratitle{Re-ranking.}
Re-ranking aims to reorder the retrieved passages as the relevant scores. Due to the small number of retrieved passages, re-ranking usually employs high-latency methods to obtain better performance, \eg cross-encoder.
Traditionally, the re-ranking task is heavily driven by manual feature engineering~\cite{guo2016deep,hui2018pacrr}. With the development of pre-trained language models~(\eg BERT), researchers use the pre-trained models to perform re-ranking tasks ~\cite{nogueira2019passage,li2020parade}.
In addition to cross-encoder, researchers also try to apply generator to re-ranking. For example, monoT5~\cite{nogueira2020doc} proposes a prompt-based method to re-rank passages with T5~\cite{raffel2020exploring} and other studies~\cite{santos2020beyond,zhuang2021deep,lesota2021modern} propose to use the log-likelihood of the query against the passage as the relevance to perform the re-ranking task. 

Recently, with the size of pre-trained models scaling up, the generative models show competitive zero-shot and few-shot ability. Researchers start to apply large generative models to zero-shot and few-shot re-ranking. For example, SGPT~\cite{muen2022sgpt} and UPR~\cite{sachan2022improving} propose to use generative models to perform zero-shot re-ranking. P$^3$ Ranker~\cite{hu2022ranker} demonstrates that generative models achieve better performance in the few-shot setting. Note that all of these works are concurrent to our work. Instead of using a query generator as a re-ranker only,  we propose to leverage the query generator as a teacher model to enhance the performance of the cross-lingual dual-encoder. In addition to the traditional knowledge distillation, we further propose a novel cost-effective alignment method to boost the dense retrieval performance in low-resource languages.

\section{Preliminaries} \label{sec:background}

In this section, we give a brief review of dense retrieval and re-ranking. The overviews of all methods are presented in Figure~\ref{fig:encoder}.

\paratitle{Dual-Encoder.} Given a query $q$ and a large corpus $C$, the retrieval task aims to find the relevant passages for the query from a large corpus. Usually, a dense retrieval model employs two dense encoders~(\eg BERT) $E_Q(\cdot)$ and $E_P(\cdot)$. They encode queries and passages into dense embeddings, respectively. Then, the model uses a similarity function, often dot-product, to perform retrieval:
\begin{equation} \label{eq:fde}
    f_{de}(q, p) = E_Q(q) \cdot E_P(p),
\end{equation}
where $q$ and $p$ denote the query and the passage, respectively. During the inference stage, we apply the passage encoder $E_P(\cdot)$ to all the passages and index them using FAISS~\cite{johnson2019billion} which is an extremely efficient, open-source library for similarity search. Then given a query $q$, we derive its embedding by $\bm{v}_q = E_Q(q)$ and retrieve the top $k$ passages with embeddings closest to $\bm{v}_q$.

\paratitle{Cross-Encoder Re-ranker.} Given a query $q$ and top $k$ retrieved passages $C$, the re-ranking task aims to reorder the passages as the relevant scores. Due to the limited size of the corpus, the re-ranking task usually employs a cross-encoder to perform interaction between words across queries and passages at the same time. These methods also introduce a special token \emph{[SEP]} to separate q and p, and then the hidden state of the \emph{[CLS] token} from the cross-encoder is fed into a fully-connected layer to output the relevant score:
\begin{equation} \label{eq:fce}
    f_{ce}(q, p) = \bm{W} \times E_C(q || p) + b,
\end{equation}
where ``$||$'' denotes concatenation with the \emph{[SEP] token}. During the inference stage, we apply the cross-encoder $E_C(\cdot)$ to all <q, p> pair and reorder the passages by the scores.

\paratitle{Query Generator Re-ranker.} Similar to cross-encoder re-ranker, query generator re-ranker also aims to reorder the passages as the relevant scores. For the query generator re-ranker, we use the log-likelyhood of the query against the passage to measure the relevance:
\begin{equation} \label{eq:fqg}
    f_{qg}(q, p) = \log P(q|p) = \sum_{t=0} \log P(q_t | q_{<t}, p),
\end{equation}
where $q_{<t}$ denotes the previous tokens before $q_t$. The rest of settings are the same as the cross-encoder re-ranker and are omitted here.

\paratitle{Training.} The goal of retrieval and re-ranking is to enlarge the relevant score between the query and the relevant passages~(\aka positive passages) and lessen the relevant score between the query and the irrelevant passages~(\aka negative passages). Let $\{q_i, p^{+}_{i}, p^{-}_{i,0}, p^{-}_{i,1}, \dots, p^{-}_{i,n}\}$ be the $i$-th training sample. It consists of a query, a positive passage, and $n$ negative passages. Then we can employ the contrastive loss function, called InfoNCE~\cite{oord2018representation}, to optimize the model:
\begin{equation} \label{eq:contrastive}
    \mathcal{L}_{R} = - \log \frac{e^{f(q_i, p^{+}_{i})}}{e^{f(q_i, p^{+}_{i})} + \sum_{j=0}^{n} e^{f(q_i, p^{-}_{i,j})}},
\end{equation}
where $f$ denotes the similarity function, \eg $f_{de}$ in Eq.~\eqref{eq:fde}, $f_{ce}$ in Eq.~\eqref{eq:fce}, or $f_{qg}$ in Eq.~\eqref{eq:fqg}.

\paratitle{Cross-lingual Retrieval.} In the cross-lingual information retrieval task, passages and queries are in different languages. In this paper, we consider the passages are in English and the queries are in non-English languages. A sample consists of three components: a query in a non-English language, a positive passage in English, and a span answer in English. Given a non-English query, the task aims to retrieve relevant passages in English to answer the query. If a retrieved passage contains the given span answer, it is regarded as a positive passage, otherwise, it is a negative passage.

\begin{figure}[t]
    \centering
    \includegraphics[width=\columnwidth]{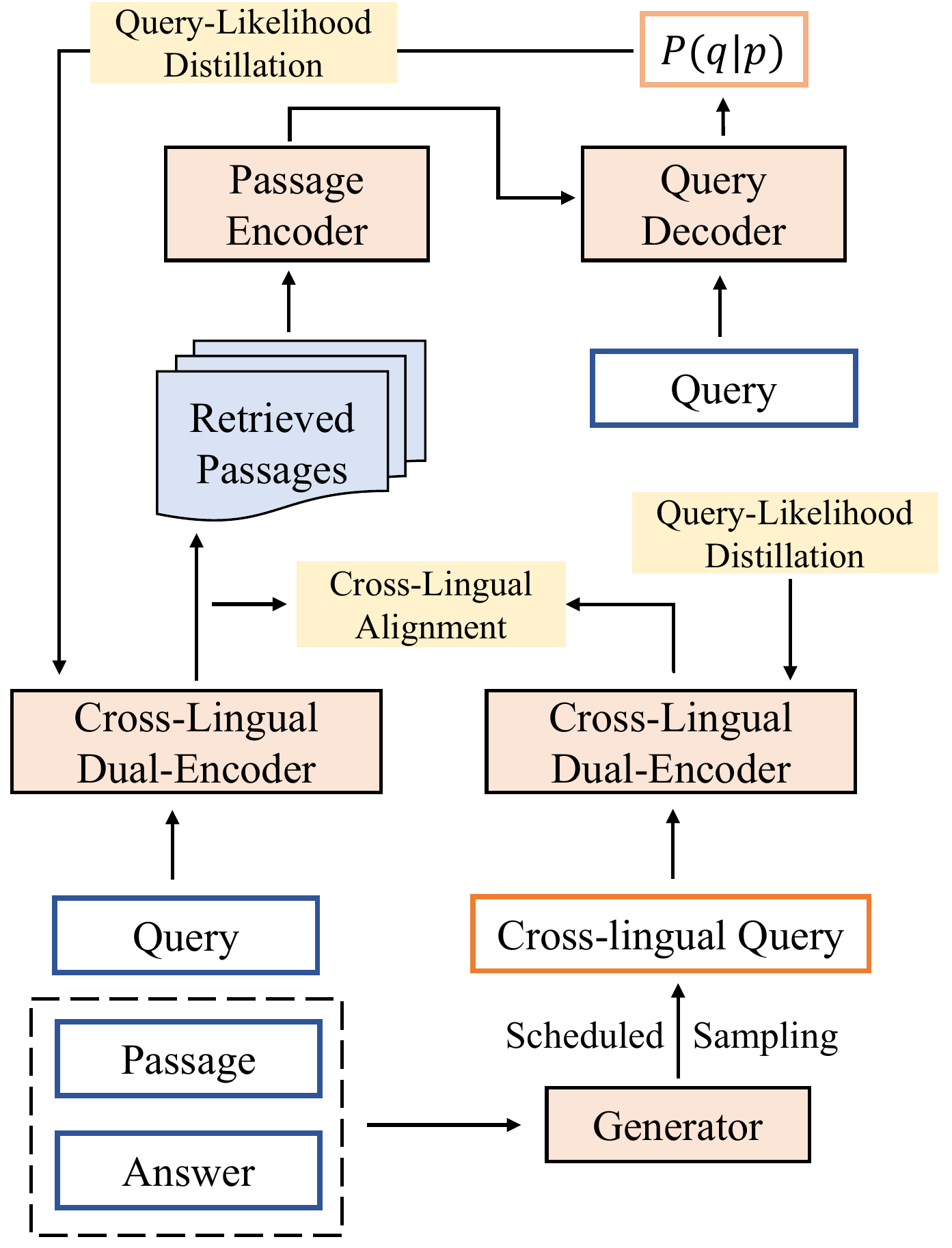}
    \caption{Overview of the proposed \name.}
    \label{fig:overview}
\end{figure}

\section{Methodology}

In this section, we present the proposed \name. The overview of the proposed method is presented in Figure~\ref{fig:overview}. We start with the training of the query generator, then present how to perform distillation and alignment training for the dual-encoder, and we finally discuss the entire training process.

\subsection{Training of Query Generator}

In our method, we employ mT5~\cite{xue2020mt5} as the query generator. The query generator has two roles: teacher and generator. As a teacher, it aims to better re-rank the candidate passages with relevance and distill the knowledge to the dual-encoder. As a generator, it aims to generate synonymous queries in different languages.

\paratitle{Input Format.}
As we employ mT5, we design a prompt template for input sentences. Considering that most passages are long, we propose introducing the span answer as input to encourage the generator to focus on the same segment and generate parallel queries in different languages. As a result, we use \emph{``generate [language] query: answer: [span answer] content: [content]''} as the template. For a specific sample, we fill the three placeholders with the language of the target query, the span answer, and the passage content, respectively.

\paratitle{Training.}
Considering the two roles of the query generator, the entire training process for the query generator contains two stages: query generation training and re-ranking training.

Firstly, we train the generator with the generation task, which takes the positive passage as input and aims to generate the query. The task can be formulated as maximizing the conditional probability:
\begin{equation}
    \begin{aligned}
        \hat{q} &= \mathop{\arg\max}\limits_{q} P(q|p, a) \\
        &= \mathop{\arg\max}\limits_{q} \prod_{t=0}P(q_t|p, a, q_{<t})
    \end{aligned},
\end{equation}
where $q_t$ is the $t$-th token of the generated query, $a$ denotes the span answer, and $q_{<t}$ represents the previous decoded tokens. Then we can employ cross-entropy loss to optimize the model:
\begin{equation} \label{eq:qg}
    \mathcal{L}_{QG} = \frac{1}{T}\sum_{t=0}^{T}-\log P(q_t|p, a, q_{<t}),
\end{equation}
where $T$ denotes the number of the query tokens.

Secondly, we train the generator with the re-ranking task, which takes a query and a passage as input and outputs the relevant score of the two sentences. The detailed training process is introduced in Section~\ref{sec:background} and is omitted here.

\subsection{Distillation for Dual-Encoder}

We then present how to distill knowledge from query generator to dual-encoder. Similar to previous methods~\cite{ren2021rocketqav2,zhang2021adver}, we employ KL-Divergence to perform distillation. Formally, given a query $q$ and a candidate passage set $C_q = \{p_i\}_{1 \le i \le n}$ which is retrieved by the dual-encoder, we compute relevant scores by query generator and dual-encoder, respectively. After that, we normalize the scores by softmax and compute the KL-Divergence as the loss:
\begin{equation} \label{eq:distillation}
    \begin{aligned}
        s_{qg}(q, p) &= \frac{\exp(f_{qg}(q, p))}{\sum_{p' \in C_q}\exp(f_{qg}(q, p'))}, \\
        s_{de}(q, p) &= \frac{\exp(f_{de}(q, p))}{\sum_{p' \in C_q}\exp(f_{de}(q, p'))}, \\
        \mathcal{L}_{D} &= \sum_{p \in C_q} s_{qg}(q, p) \frac{s_{qg}(q, p)}{s_{de}(q, p)},
    \end{aligned}
\end{equation}
where $f_{qg}$ and $f_{de}$ denote the relevant score given by the query generator and the dual-encoder which are presented in Section~\ref{sec:background}.

\subsection{Alignment for Dual-Encoder}

Alignment is a common topic in the cross-lingual setting, which can help the model better handle sentences in different languages. Previous works~\cite{zheng2021consistency,yang2022enhancing} usually use parallel data or translated data to perform alignment training among different languages. Here, we propose a novel method to align queries in different languages for cross-lingual retrieval, which does not need any parallel data. The core idea of our method is to leverage the query generator to generate synonymous queries in other languages to form parallel cases.

\paratitle{Generation.} For each case in the training set, we generate a query in each target language~(\aka if there are seven target languages, we generate seven queries for the case). Then we use the confidence of the generator to filter the generated queries. Specially, we set filter thresholds to accept 50\% of generated queries. 

\paratitle{Scheduled Sampling.} In this work, we select a generated query to form a pair-wise case with the source query. Considering the semantics of generated queries, we carefully design a scheduled sampling method to replace the random sampling. For a generated query $q'$, we first use the dual-encoder to retrieve passages for the source query $q$ and generated query $q'$, respectively, namely $C_{q}$ and $C_{q}'$. Then we calculate a coefficient for the generated query $q'$ as
\begin{equation} \label{eq:thresh}
    \begin{aligned}
        c' &= \frac{|C_q \cap C_q'|}{\max(|C_q|, |C_q'|)}, \\
        c' &= \begin{cases}
            c' & \text{ if } c' \ge T,\\
            0 & \text{ if } c' < T,
        \end{cases}
    \end{aligned}
\end{equation}
where threshold $T$ is a hyper-parameter and $|\cdot|$ denotes the size of the set. The basic idea is that the larger the union of retrieved passages, the more likely the queries are to be synonymous.
When sampling the generated query, we first calculate coefficients $\{c_1', \dots, c_m'\}$ for all generated queries $\{q_1', \dots, q_m'\}$, then normalize them as the final sampling probability $p$:
\begin{equation}
    p_i = \frac{c_i'}{\sum_{j=0}^{m}c_i'},
\end{equation}
where $m$ denotes the number of generated queries. During the training stage, for each training case, we sample a generated query to form the pair-case with the source query $q$ based on the probabilities.

\paratitle{Alignment Training.}
After sampling a generated query, we present the how to align the source query and the generated query. Different to previous works~\cite{zheng2021consistency}, we employ asymmetric KL-Divergence rather than symmetric KL-Divergence due to the different quality of the source query and the generated query:
\begin{equation} \label{eq:alignment}
    \mathcal{L}_{A} = \sum_{p \in C_q \cup C_q'} c' s_{de}(q, p) \frac{s_{de}(q, p)}{s_{de}(q', p)},
\end{equation}
where $q$ denotes the query, $C_q$ denotes the set of retrieved passages, superscript ``$\prime$'' denotes the generated case, and $c'$ is the coefficient of the generated query. Note that $s_{de}$ in Eq.~\eqref{eq:alignment} are normalized across $C_q \cup C_q'$ instead of $C_q$ or $C_q'$ in Eq.~\eqref{eq:distillation}.

\subsection{Training of Dual-Encoder}

As shown in Figure~\ref{fig:overview}, we combine the distillation loss and the alignment loss as final loss:
\begin{equation} \label{eq:final}
    \mathcal{L} = \mathcal{L}_{D} + \mathcal{L}_{D}' + \alpha \times \mathcal{L}_{A},
\end{equation}
where $\mathcal{L}_{D}$ denotes the distillation loss for the source queries, $\mathcal{L}_{D}'$ denotes the distillation loss for the generated queries, $\mathcal{L}_{A}$ denotes the alignment loss, and $\alpha$ is a hyper-parameter to balance the loss. 

Based on the training method of dual-encoder and query generator, we conduct an iterative procedure to improve the performance. We present the entire training procedure in Algorithm~\ref{alg:training}.

\begin{algorithm}[t]
    \caption{The training algorithm.}
    \label{alg:training} 
    \LinesNumbered
    \KwIn{Dual-Encoder $R$, Query Generator $G$, Corpus $C$, and Training Set $D$.}
    
    Initialize $R$ and $G$ with pre-trained model\;
    Train the warm-up $R$ with Eq.~\eqref{eq:contrastive} on $D$\;
    Train the warm-up $G$ with Eq.~\eqref{eq:qg} on $D$\;
    Generate queries for each sample in $D$\;
    Build ANN index for $R$\;
    Retrieve relevant passages on corpus $C$\;
    Fine-tune the $G$ with Eq.~\eqref{eq:contrastive} on $D$ and retrieved negative passages.
    
    \While{models has not converged}{
        Fine-tune the $R$ with Eq.~\eqref{eq:final} on $D$ and retrieved passages\;
        Refresh ANN index for $R$\;
        Retrieve relevant passages on corpus $C$\;
        Fine-tune the $G$ with Eq.~\eqref{eq:contrastive} on $D$ and retrieved negative passages.
    }
\end{algorithm}
\section{Experiments} \label{sec:exp}

In this section, we construct experiments to demonstrate the effectiveness of our method.

\subsection{Experimental Setup}

\paratitle{Datasets.} We evaluate the proposed method on two public cross-lingual retrieval datasets: XOR-Retrieve~\cite{asai2020xor} and MKQA~\cite{longpre2020mkqa}. The detailed descriptions of the two datasets are presented in Appendix~\ref{sec:dataset}.

\paratitle{Evaluation Metrics.} Following previous works \cite{asai2020xor, soro2022ask}, we use R@2kt and R@5kt as evaluation metrics for the XOR-Retrieve dataset and R@2kt as evaluation metrics for the MKQA dataset. The metrics measure the proportion of queries to which the top k retrieved tokens contain the span answer, which is fairer with different passage sizes.

\paratitle{Implementation Details.} For the warm-up training stage, we follow XOR-Retrieve to first train the model on NQ~\cite{kwiatkowski2019natural} data and then fine-tune the model with XOR-Retrieve data. For the iteratively training stage, we generate seven queries for each case~(because the XOR-Retrieve data contains seven languages). We set the number of retrieved passages as 100, the number of iterations as $5$, threshold $T$ in Eq.~\eqref{eq:thresh} as 0.3 and coefficient $\alpha$ in Eq.~\eqref{eq:final} as 0.5. The detailed hyper-parameters are shown in Appendix~\ref{sec:param}. And we conduct more experiments to analyze the parameter sensitivity in Appendix~\ref{sec:sens}.

All the experiments run on 8 NVIDIA Tesla A100 GPUs. The implementation code is based on HuggingFace Transformers~\cite{wolf2020transformers}. For the dual-encoder, we use XLM-R Base~\cite{conneau2019unsupervised} as the pre-trained model and use the average hidden states of all tokens to represent the sentence. For the query generator, we leverage mT5 Base~\cite{xue2020mt5} as the pre-trained model, which has almost the same number of parameters as a large cross-encoder.

\begin{table*}[t] \footnotesize
\centering
\setlength\tabcolsep{4pt}
\caption{Comparison results on XOR-Retrieve dev set. The best results are in bold. ``$\ast$'' denotes the results are copied from the source paper. Results unavailable are left blank.}
\begin{tabular}{l|ccccccc|c|ccccccc|c} \toprule
    \multirow{2.5}{*}{Methods} & \multicolumn{8}{c|}{R@2kt} & \multicolumn{8}{c}{R@5kt} \\ \cmidrule(lr){2-17}
    ~ & Ar & Bn & Fi & Ja & Ko & Ru & Te & Avg & Ar & Bn & Fi & Ja & Ko & Ru & Te & Avg \\ \midrule
    mDPR$^{\ast}$ & 38.8 & 48.4 & 52.5 & 26.6 & 44.2 & 33.3 & 39.9 & 40.5 & 48.9 & 60.2 & 59.2 & 34.9 & 49.8 & 43.0 & 55.5 & 50.2 \\
    DPR + MT$^{\ast}$ & 43.4 & 53.9 & 55.1 & 40.2 & 50.5 & 30.8 & 20.2 & 42.0 & 52.4 & 62.8 & 61.8 & 48.1 & 58.6 & 37.8 & 32.4 & 50.6 \\ \midrule
    Sentri$^{\ast}$ & 47.6 & 48.1 & 53.1 & 46.6 & 49.6 & 44.3 & 67.9 & 51.0 & 56.8 & 62.2 & 65.5 & 53.2 & 55.5 & 52.3 & 80.3 & 60.8 \\
    ~~~ w/ Bi-Encoder$^{\ast}$ & 47.8 & 39.1 & 48.9 & 51.2 & 40.2 & 41.2 & 49.4 & 45.4 & 55.1 & 43.3 & 59.5 & 59.4 & 51.2 & 52.0 & 56.9 & 53.9 \\ \midrule
    DR.DECR$^{\ast}$ & - & - & - & - & - & - & - & 66.0 & 70.2 & \textbf{85.9} & 69.4 & 65.1 & 68.8 & 68.8 & 83.2 & 73.1 \\
    ~~~ w/o KD$_{XOR}^{\ast}$ & - & - & - & - & - & - & - & 60.6 & - & - & - & - & - & - & - & 68.6 \\
    ~~~ w/o KD$_{PC}^{\ast}$ & - & - & - & - & - & - & - & 56.6 & - & - & - & - & - & - & - & 63.6 \\\midrule
    \name & 52.8 & 70.1 & 62.2 & 54.8 & 62.8 & 57.8 & 70.6 & 61.3 & 63.8 & 78.0 & 65.3 & 63.5 & 69.8 & 67.1 & 74.8 & 68.9 \\ 
    \name w/ LaBSE & \textbf{67.3} & \textbf{78.9} & \textbf{65.9} & \textbf{59.8} & \textbf{66.3} & \textbf{63.7} & \textbf{80.7} & \textbf{68.9} & \textbf{72.2} & 83.2 & \textbf{69.7} & \textbf{68.0} & \textbf{70.9} & \textbf{71.7} & \textbf{84.9} & \textbf{74.4} \\ \bottomrule
\end{tabular}
\label{tab:xor-dev}
\end{table*}

\begin{table}[t] \small
\centering

\caption{Comparison results on XOR-Retrieve test set. }
\begin{tabular}{l|cc} \toprule
    Methods & R@2kt & R@5kt \\ \midrule
    GAAMA & 52.8 & 59.9 \\
    Sentri & 52.7 & 61.0 \\
    CCP & 54.8 & 63.0 \\
    Sentri 2.0 & 58.5 & 64.6 \\
    DR.DECR & 63.0 & 70.3 \\
    \name w/ LaBSE & \textbf{65.6} & \textbf{72.0} \\ \bottomrule
\end{tabular}
\label{tab:xor-test}
\end{table}

\subsection{Results}

\paratitle{Baselines.} We compare the proposed \name with previous state-of-the-art methods, including mDPR, DPR+MT~\cite{asai2020xor}, Sentri~\cite{soro2022ask}, DR.DECR~\cite{li2021learning}. Note that Sentri introduces a shared encoder with large size, DR.DECR introduces parallel queries and parallel corpus, but our method only utilizes an encoder with base size, XOR-Retrieve and NQ training data. For more fairly comparison, we also report their ablation results. Here, ``Bi-Encoder'' denotes two unshared encoders with base size. ``KD$_{XOR}$'' denotes a distillation method which introduces synonymous English queries. ``KD$_{PC}$'' denotes a distillation method which introduces parallel corpus. 
In addition, we also employ LaBSE base~\cite{feng2022language} to evaluate the proposed \name with parallel corpus, which is a state-of-the-art model pre-trained with parallel corpus.

\paratitle{XOR-Retrieve.}
Table~\ref{tab:xor-dev} shows the results on XOR-Retrieve dev set. The proposed \name outperforms mDPR, DPR+MT, and Sentri with a clear edge in almost all languages. Although \name does not introduce any parallel corpus, it also outperforms DR.DECR w/o KD$_{XOR}$. Finally, \name based on LaBSE outperforms all baselines, especially DR.DECR w/o KD$_{XOR}$, and even outperforms DR.DECR which utilizes both parallel queries and parallel corpus. Note that knowledge distillation with parallel corpus in DR.DECR is designed for cross-lingual dense retrieval, but LaBSE is a general pre-trained model for all cross-lingual tasks. These results show the effectiveness of the proposed \name. Our method combines two methods in dense retrieval and cross-lingual tasks, namely distillation and alignment. We further analyze the contribution of each component in Section~\ref{sec:abla}.

In addition, we show the results on XOR-Retrieve test set in Table~\ref{tab:xor-test}, which is copied from the leaderboard\footnote{\url{https://nlp.cs.washington.edu/xorqa}} on June 15, 2022. As we can see, our method achieves the top position on the leaderboard of XOR-Retrieve.

\begin{table}[t] \small
\centering

\caption{Average performance of 20 unseen languages in MKQA test set. ``$\ast$'' denotes the results are copied from the Sentri paper.}
\begin{tabular}{l|c} \toprule
    Methods & R@2kt \\ \midrule
    CORA$^{\ast}$ & 41.1 \\
    BM25 + MT$^{\ast}$ & 42.0 \\
    Sentri$^{\ast}$ & 53.3 \\
    ~~~ w/ Bi-Encoder$^{\ast}$ & 45.3 \\ \midrule
    \name & 53.4 \\
    \name w/ LaBSE & \textbf{60.3} \\ \bottomrule
\end{tabular}
\label{tab:mkqa}
\end{table}

\paratitle{MKQA.}
Furthermore, we evaluate the zero-shot performance of our method on the MKQA test set. Following previous works~\cite{soro2022ask}, we directly evaluate the dual-encoder training on XOR-Retrieve data and report the performance of unseen languages on MKQA. As shown in Table~\ref{tab:mkqa}, our method outperforms all baselines and even performs better than Sentri. Note that Sentri uses a shared encoder with large size. The comparison between Sentri and Sentri w/ Bi-Encoder shows that the large encoder has better transfer ability. Finally, the proposed \name w/ LaBSE outperforms all baselines with a clear edge.  It shows the better transfer ability of our methods.

\begin{table}[t] \small
\centering

\caption{Ablation results on XOR-Retrieve dev set. }
\begin{tabular}{l|cc} \toprule
    Methods & R@2kt & R@5kt \\ \midrule
    \name & \textbf{61.3} & \textbf{68.9} \\
    w/o Sampling & 59.5 & 67.5 \\
    w/o Alignment & 59.9 & 67.1 \\
    w/o Generation & 58.8 & 65.9 \\ 
    w/o All & 41.5 & 53.4 \\ \bottomrule
\end{tabular}
\label{tab:ablation}
\end{table}

\subsection{Methods Analysis} \label{sec:abla}

\paratitle{Ablation Study.} 
Here, we check how each component contributes to the final performance. We construct the ablation experiments on XOR-Retrieve data. We prepare four variants of our method: 
(1) \uline{w/o Sampling} denotes without the scheduled sampling but keep the threshold $T$ for $c'$, \aka if $c' \ge T$, then $c' = 1$, otherwise $c' = 0$;
(2) \uline{w/o Alignment} denotes without $\mathcal{L}_A$ in Eq.~\eqref{eq:final};
(3) \uline{w/o Generation} denotes without $\mathcal{L}_D'$ and $\mathcal{L}_A$ in Eq.~\eqref{eq:final};
(4) \uline{w/o All} denotes without the enhanced training, \aka the warm-up dual-encoder.

Table~\ref{tab:ablation} presents all comparison results of the four variants. 
As we can see, the performance rank of R@5kt can be given as: w/o All < w/o Generation < w/o Alignment < w/o Sampling < \name. These results indicate that all components are essential to improve performance. And we can find the margin between w/o Alignment and w/o Sampling is small, it denotes that the generated queries are noisy and demonstrate the effectiveness of our schedule sampling strategy.

\paratitle{Effect of Alignment.}
As we mentioned in Section~\ref{sec:intro}, the alignment established in the pre-training stage may be damaged without any constraint in the fine-tuning stage. Here, we construct experiments on both XLM-R and LaBSE to analyze the effectiveness of the proposed alignment training. As shown in Table~\ref{tab:align}, the proposed alignment training is effective based on the two models. It indicates that the alignment constraint in the fine-tuning stage is effective for models which pre-trained with parallel corpus. And we find that the gains of alignment training based on XLM-R are larger than LaBSE, which shows that the alignment constraint is more effective for models which do not pre-trained with parallel corpus.

\begin{table}[t] \small
\centering
\setlength\tabcolsep{4pt}

\caption{Effect of alignment based on different pre-trained languages models.}
\begin{tabular}{l|cc|cc} \toprule
    \multirow{2.5}{*}{Methods} &\multicolumn{2}{c|}{XLM-R} & \multicolumn{2}{c}{LaBSE} \\ \cmidrule(lr){2-5}
    ~ & R@2kt & R@5kt & R@2kt & R@5kt \\ \midrule
    \name & \textbf{61.3} & \textbf{68.9} & \textbf{68.9} & \textbf{74.4} \\
    ~~~ w/o Alignment & 59.9 & 67.1 & 67.9 & 73.8 \\ \bottomrule
\end{tabular}
\label{tab:align}
\end{table}

\begin{figure}[t]
    \centering
    \includegraphics[width=0.78\columnwidth]{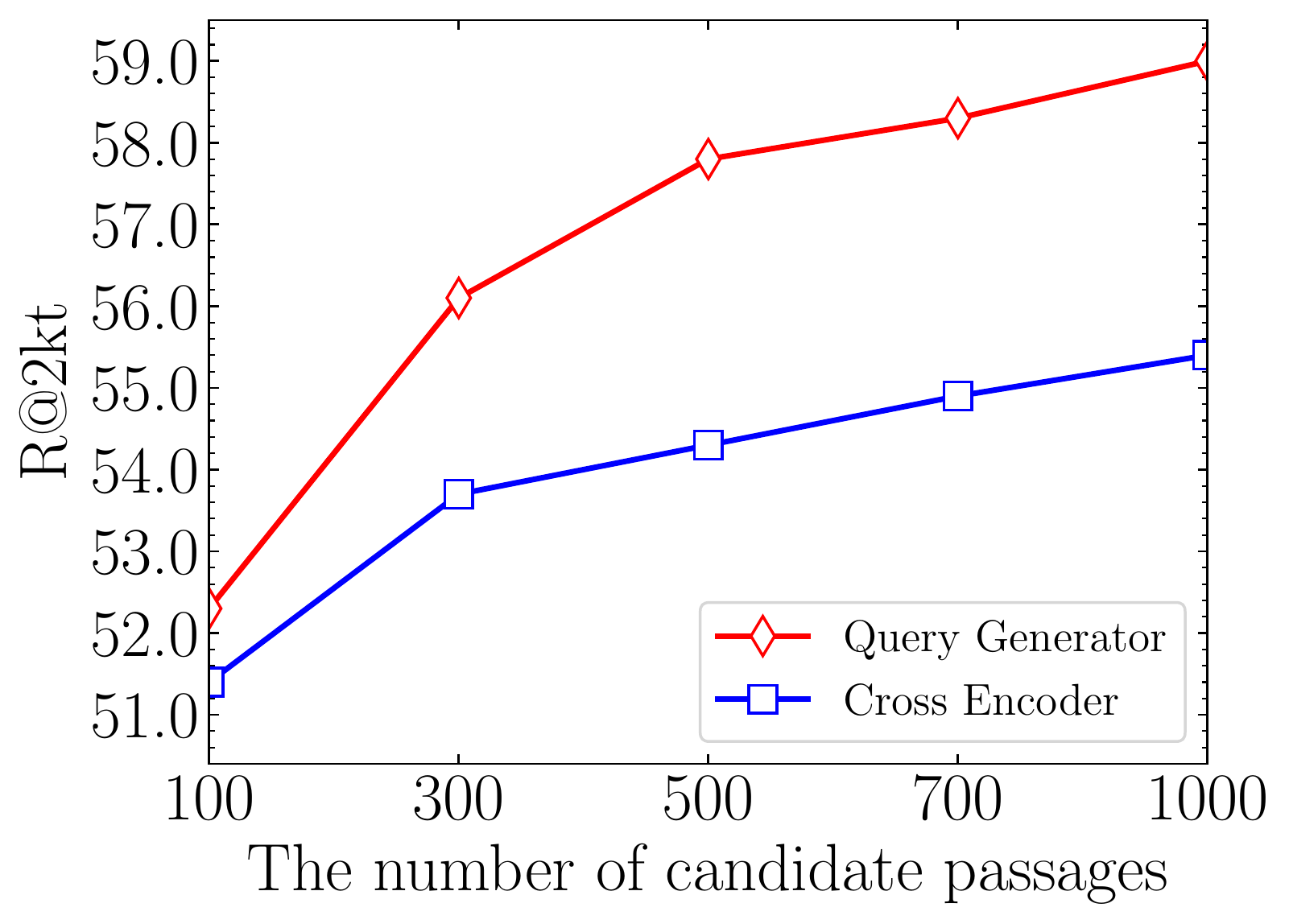}
    \caption{Re-ranking performance of cross-encoder and query generator on XOR-Retrieve dev set with different numbers of candidate passages.}
    \label{fig:depth}
\end{figure}

\begin{figure}[t]
    \centering
    \includegraphics[width=0.75\columnwidth]{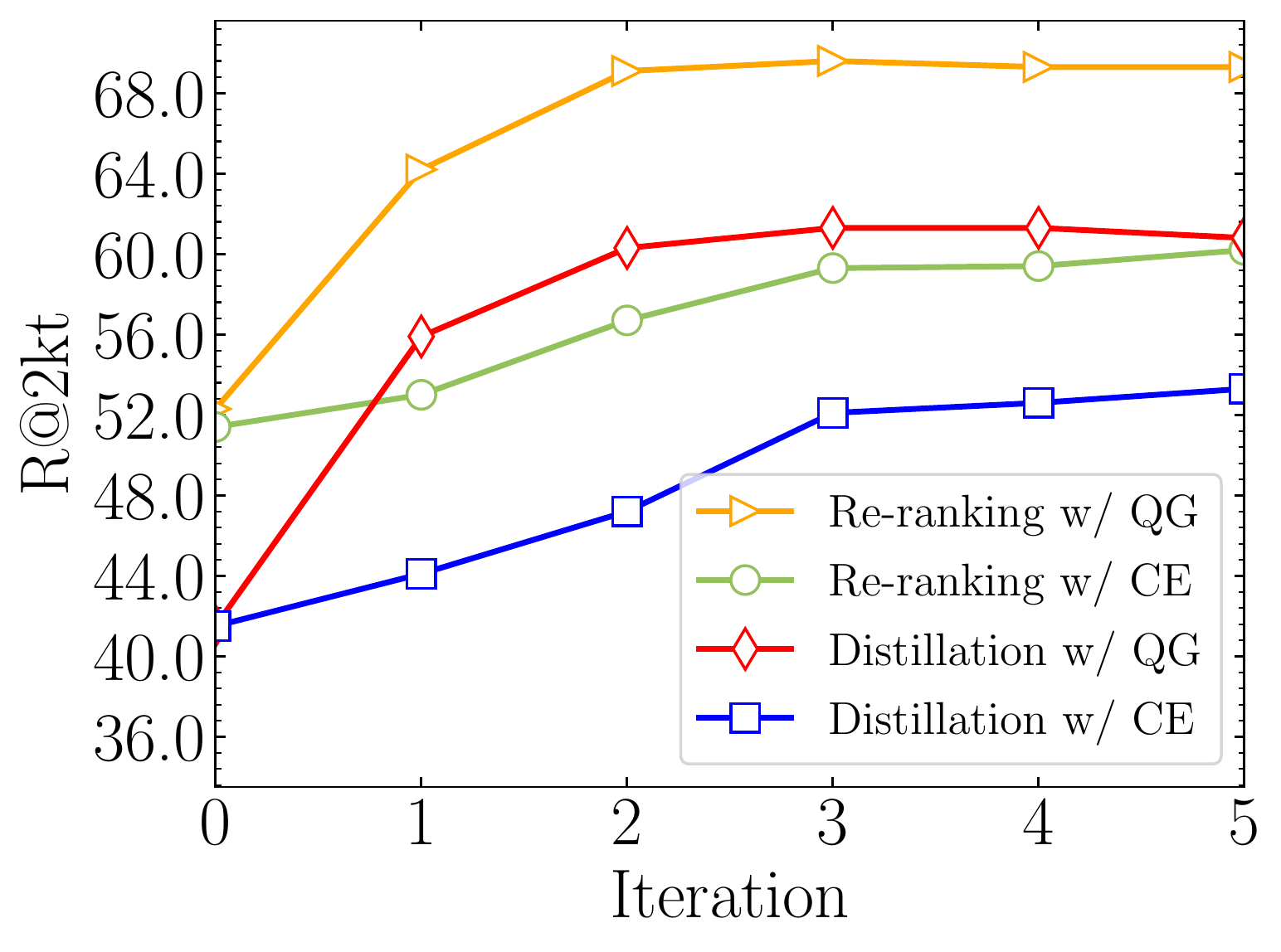}
    \caption{The changes of R@2kt during the iteratively training on XOR-Retrieve dev set. Here, ``QG'' denotes Query Generator and ``CE'' denotes Cross Encoder.}
    \label{fig:iteration}
\end{figure}

\paratitle{Cross-Encoder versus Query Generator.} Here, we analyze the re-ranking ability of cross-encoder and query generator. Here, we use the warm-up dual-encoder to retrieve passages, vary the number of candidate passages, and then evaluate the re-ranked result. As shown in Figure~\ref{fig:depth}, when we use the top-100 candidate passages, the performance of the cross-encoder and generator is almost the same. But as the number of candidate passages increases, especially when it surpasses 500, the gap between the performance of the cross-encoder and query generator gradually becomes larger. It shows low generalization performance of the cross-encoder when there are not enough training samples.

\paratitle{Visualization of the Training Procedure.} We visualize the performance changes of R@2kt during the training of both dual-encoder and query generator re-ranker which re-ranks the retrieved top-100 passages. We also incorporate a cross-encoder~(initialized with XLM-R Large) to perform distillation and re-ranking for comparison. As shown in Figure~\ref{fig:iteration}, the R@2kt of all models gradually increases as the iteration increases. While the training advances closer to convergence, the improvement gradually slows down. In the end, the performance of the dual-encoder is improved by approximately 17\%, and the performance of the query generator is improved by approximately 20\%. 
Finally, comparing the performance of the cross-encoder and the query generator, we can find that there are approximately 6\% gaps for both teachers and students. It shows the effectiveness of our method. 

\section{Conclusion}

In this paper, we showed that the cross-encoder performs poorly when there are not sufficient training samples which are hard to obtain in the cross-lingual setting. Then we proposed a novel method that utilizes the query generator to improve the dual-encoder.
We firstly proposed to use a query generator as the teacher.
After that, we proposed a novel alignment method for cross-lingual retrieval which does not need any parallel corpus. 
Extensive experimental results show that the proposed method outperforms the baselines and significantly improves the state-of-the-art performance. 
Currently, our method depends on training data in all target languages. As future work, we will investigate how to perform the proposed method for zero-shot cross-lingual dense retrieval.

\section{Limitations}

The limitations are summarized as follows.

\begin{itemize}
    \item The method depends on training data in all target languages. Intuitively, the method can be directly applied to the zero-shot cross-lingual dense retrieval if we only take the passage as input for the query generator, but the query generator performs poorly in the zero-shot setting. As future work, novel pre-training tasks for cross-lingual generation can be considered.
    \item The method does not investigate how to effectively train the query generator for the re-ranking task, just directly applies the training method for the cross-encoder re-ranker. We believe the potential of query generators for re-ranking is strong and designing a special re-ranking training method for query generators such as token-level supervision may be interesting for future work.
    \item The method requires large GPU sources. The final model approximately costs 12 hours on 8 NVIDIA Tesla A100 GPUs. Although researchers who do not have enough GPU sources can use the ``gradient accumulation'' technique to reduce GPU memory consumption, they also need to pay more time.
    \item This work does not consider the inconsistency between different countries~(\eg law and religion), which leads to inconsistent positive passages for synonymous queries in different languages~(\eg the legal age of marriage varies from country to country). Because we find that most of the queries in XOR-Retrieve contain the target country such as \emph{``Mikä on yleisin kissa laji Suomessa?''~(translation: What is the most common cat breed in Finland?)}. 
\end{itemize}

\bibliography{references}
\bibliographystyle{acl_natbib}

\clearpage

\section*{Appendix}
\appendix

\section{Dataset} \label{sec:dataset}

\paratitle{XOR-Retrieve.} XOR-Retrieve dataset is a cross-lingual retrieval dataset which aims to retrieve relevant passages from the English corpus for non-English queries. XOR-Retrieve data contains queries in seven typologically diverse languages: Arabic~(Ar), Bengali~(Bn), Finnish~(Fi), Japanese~(Ja), Korean~(Ko), Russian~(Ru), and Telugu~(Te). The statistics are presented in Table~\ref{tab:stat}.

\begin{table}[h] \small
\centering
\caption{Data statistics for XOR-Retrieve.}
\begin{tabular}{c|rrr} \toprule
    ~  & Train & Dev & Test \\ \midrule
    Ar & 2,574 & 350 & 137 \\
    Bn & 2,582 & 312 & 128 \\
    Fi & 2,088 & 360 & 530 \\
    Ja & 2,288 & 296 & 449 \\
    Ko & 2,469 & 299 & 646 \\
    Ru & 1,941 & 366 & 235 \\
    Te & 1,308 & 238 & 374 \\ \midrule
    Corpus size & \multicolumn{3}{c}{18,003,200} \\ \bottomrule
\end{tabular}
\label{tab:stat}
\end{table}

\paratitle{MKQA.} MKQA dataset is a translated dataset of 10,000 query-answer pairs from NQ to 26 different languages, and the dataset is only used for evaluation. In our experiments, since we measure the R@2kt score, we filter the samples which do not have span answers. Then, we get 6,620 parallel queries in each language. Finally, we directly evaluate the dual-encoder trained on XOR-Retrieve data and use the same corpus with XOR-Retrieve in the experiments. Note that Arabic~(Ar), English~(En), Finnish~(Fi), Japanese~(Ja), Korean~(Ko), and Russian~(Ru) are seen in the training stage and we only report the performance of the rest 20 languages. As a result, it has a total of 132,400 samples.

\section{Efficiency Report}

We list the time cost of training and inference in Table~\ref{tab:efficiency} which is made with 8 NVIDIA A100 GPUs.

\begin{table}[b] \small
\centering
\caption{Efficiency Report.}
\begin{tabular}{l|l|c} \toprule
     \multirow{5}{*}{Training} & Warm-up & 3h \\
     ~ & Per Iteration of Dual-Encoder & 1h \\
     ~ & Per Iteration of Generator & 0.3h \\
     ~ & Index Refresh & 0.35h \\
     ~ & Overall & 11.5h \\ \midrule
     \multirow{3}{*}{Inference} & Build Index & 0.35h \\
     ~ & Query Encoding & 40ns \\
     ~ & Dense Retrieval & 2ms \\ \midrule
\end{tabular}
\label{tab:efficiency}
\end{table}

\section{Hyper-parameters} \label{sec:param}

We present all hyper-parameters in Table~\ref{tab:params}. 

\begin{table}[t] \small
\centering
\caption{Hyper-parameters.}
\begin{tabular}{c|l|c} \toprule
     ~ & Parameters & Value \\ \midrule
     ~ & Max Query Length & 32 \\
     ~ & Max Passage Length & 128 \\ \midrule
     \multirow{8}{*}{\shortstack{Training \\ Warm-up \\ Dual-Encoder}} & Learning Rate & 1e-5 \\
     ~ & Batch Size & 128 \\
     ~ & Negative Size & 255 \\
     ~ & Optimizer & AdamW \\ 
     ~ & Scheduler & Linear \\
     ~ & Warmup Proportion & 0.1 \\
     ~ & Training Steps on NQ & 18400 \\
     ~ & Training Steps on XOR & 2000 \\\midrule
     \multirow{6}{*}{\shortstack{Training \\ Warm-up \\ Generator \\ on QG}} & Learning Rate & 1e-4 \\
     ~ & Batch Size & 64 \\
     ~ & Optimizer & AdamW \\ 
     ~ & Scheduler & Linear \\
     ~ & Warmup Proportion & 0.1 \\
     ~ & Training Steps & 5000 \\ \midrule
     \multirow{7}{*}{\shortstack{Training \\ Warm-up \\ Generator \\ on Re-ranking}} & Learning Rate & 1e-5 \\
     ~ & Batch Size & 32 \\
     ~ & Negative Size & 15 \\
     ~ & Optimizer & AdamW \\ 
     ~ & Scheduler & Linear \\
     ~ & Warmup Proportion & 0.1 \\
     ~ & Training Steps & 1000 \\ \midrule
     \multirow{10}{*}{\shortstack{Iteraively \\ Training of \\ Dual-Encoder}} & Learning Rate & 1e-5 \\
     ~ & Batch Size & 64 \\
     ~ & Candidate Size & 32 \\
     ~ & Optimizer & AdamW \\ 
     ~ & Scheduler & Linear \\
     ~ & Warmup Proportion & 0.1 \\
     ~ & Training Steps & 3000 \\
     ~ & Threshold $T$ in Eq.~\eqref{eq:thresh} & 0.3 \\
     ~ & Coefficient $\alpha$ in Eq.~\eqref{eq:final} & 0.5 \\
     ~ & \# of iterations & 5 \\ \midrule
     \multirow{8}{*}{\shortstack{Iteraively \\ Training of \\ Generator}} & Learning Rate & 1e-5 \\
     ~ & Batch Size & 32 \\
     ~ & Negative Size & 15 \\
     ~ & Optimizer & AdamW \\ 
     ~ & Scheduler & Linear \\
     ~ & Warmup Proportion & 0.1 \\
     ~ & Training Steps & 500 \\
     ~ & \# of iterations & 5 \\ \bottomrule
\end{tabular}
\label{tab:params}
\end{table}

\begin{figure}[t]
    \centering
    \subfigure[Threshold~($T$).]{
        \includegraphics[width=0.75\columnwidth]{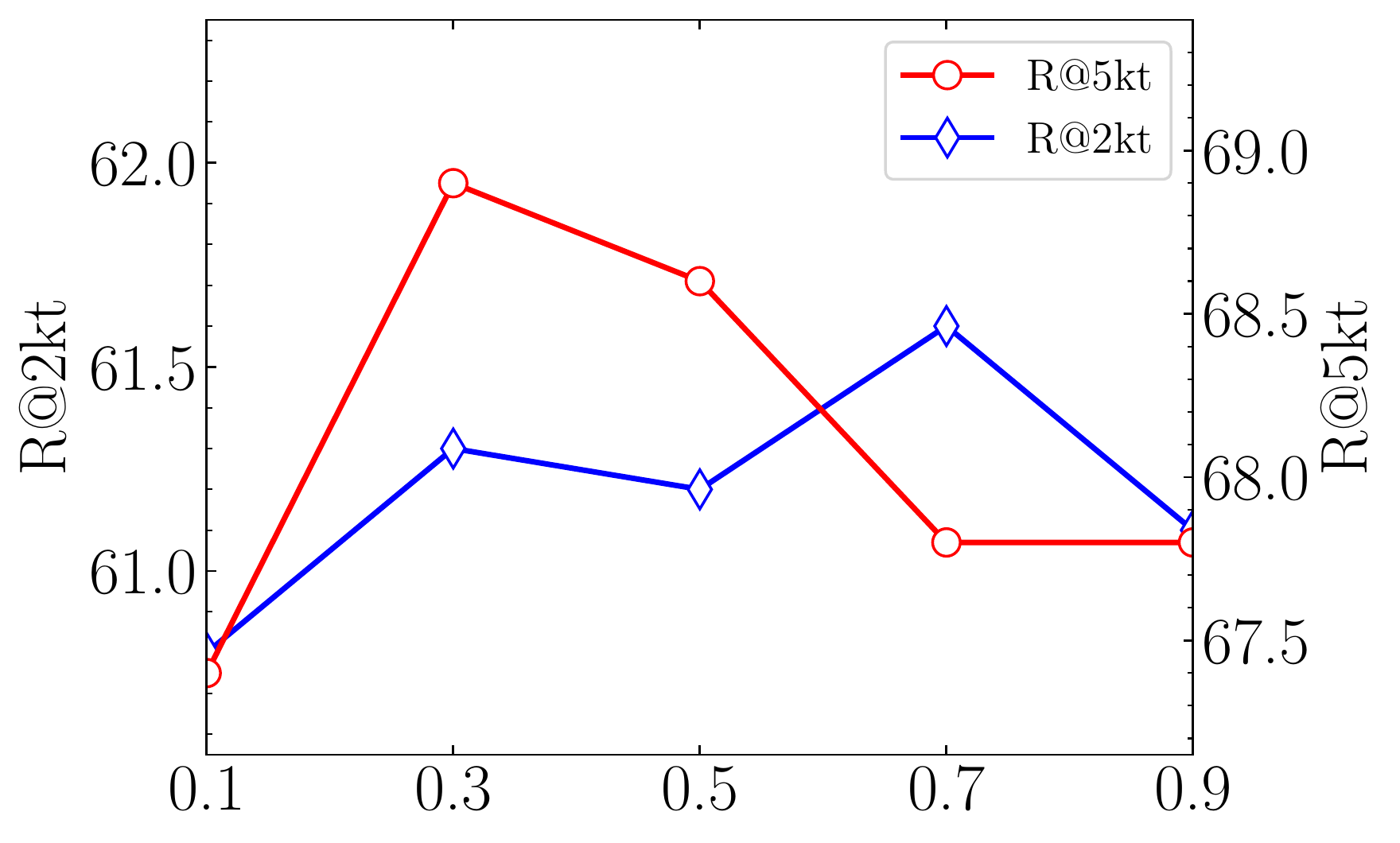}
    }
    \subfigure[Coefficient~($\alpha$).]{
        \includegraphics[width=0.75\columnwidth]{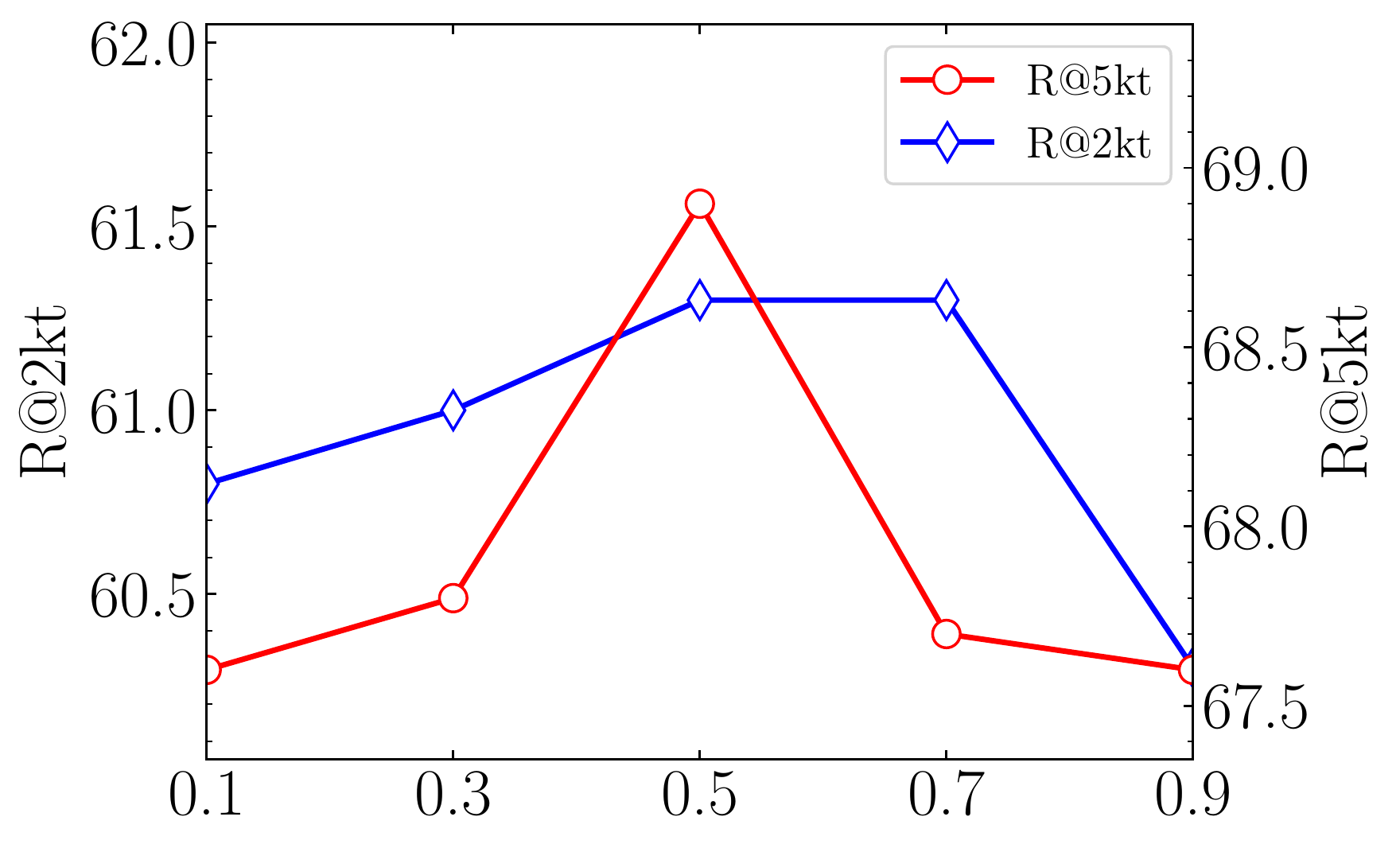}
    }
    \caption{Parameter sensitivity.}
    \label{fig:sensitivity}
\end{figure}

\section{Additional Experiments} \label{sec:sens}

\subsection{Parameter Sensitivity}

In this section, we tune the parameters of the proposed method to analyze parameter sensitivity. We vary both the threshold $T$~(Eq.~\eqref{eq:thresh}) and the coefficient $\alpha$~(Eq.~\eqref{eq:final}) in the set $\{0.1,$ $0.3,$ $0.5,$ $0.7,$ $0.9\}$. We report the tuning results with both R@2kt and R@5kt on the XOR-Retrieve dev set in Figure~\ref{fig:sensitivity}. As we can see, $T = 0.3$ and $\alpha = 0.5$ lead to the optimal R@5kt which is the ordering basis on the leaderboard.

In addition, we find that the optimal R@2kt and R@5kt are led by different thresholds $T$. Because the two metrics have different sensitivities to data quality, low-quality data is helpful to R@5kt but harmful to R@2kt. As a result, a small threshold $T$ leads to more low-quality alignment training data and further leads to higher R@5kt but lower R@2kt. On the contrary, the optimal R@2kt and R@5kt are led by the same coefficient $\alpha$.

Overall, our model is relatively stable when varying the two parameters, and consistently better than Sentri and Dr.DECR w/o KD$_{PC}$.

\subsection{Effect of The Number of Candidates}

\begin{figure}[t]
    \centering
    \includegraphics[width=0.66\columnwidth]{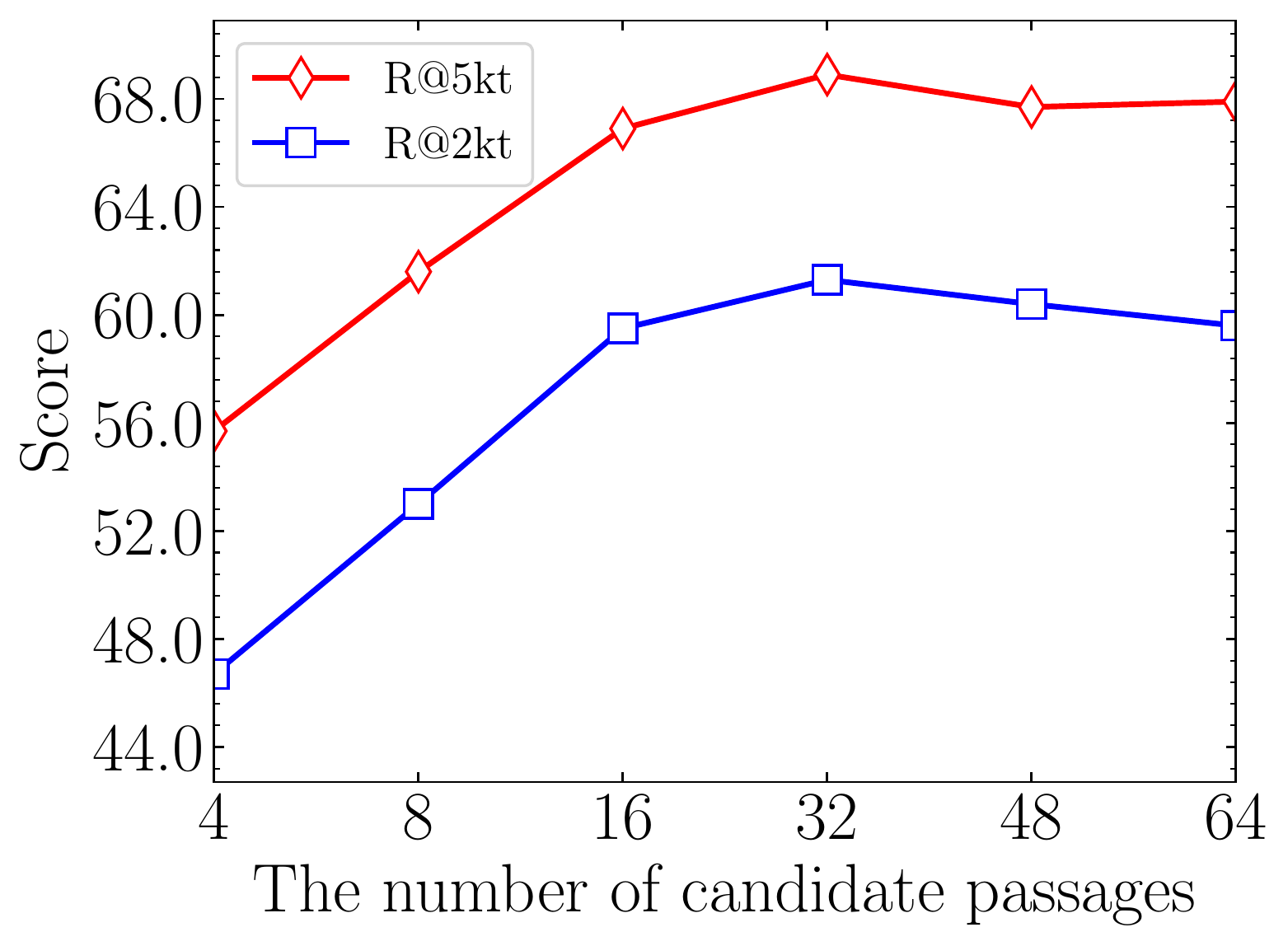}
    \caption{Effect of the number of candidate passages.}
    \label{fig:candidate}
\end{figure}

Here, we investigate the effect of the number of candidates which is demonstrated to have a significant effect on the final performance. As shown in Figure~\ref{fig:candidate}, a large number of candidates leads to better performance. And when the number surpasses 32, the improvement gradually slows down. The results indicate that 32 candidates can better represent the whole corpus.

\subsection{Effect of Model Size}

\begin{table}[t] \small
\centering
\caption{Effect of model size.}
\begin{tabular}{c|cc|cc} \toprule
    \multirow{2.5}{*}{~} &\multicolumn{2}{c|}{XLM-R Base} & \multicolumn{2}{c}{XLM-R Large} \\ \cmidrule(lr){2-5}
    ~ & R@2kt & R@5kt & R@2kt & R@5kt \\ \midrule
    Ar & 52.8 & 63.8 & \textbf{64.4} & \textbf{72.2} \\
    Bn & 70.1 & 78.0 & \textbf{77.0} & \textbf{80.9} \\
    Fi & 62.2 & 65.3 & \textbf{67.0} & \textbf{70.1} \\
    Ja & 54.8 & 63.5 & \textbf{56.4} & \textbf{67.2} \\
    Ko & 62.8 & 69.8 & \textbf{67.4} & \textbf{73.3} \\
    Ru & 57.8 & 67.1 & \textbf{66.7} & \textbf{72.2} \\
    Te & 70.6 & 74.8 & \textbf{79.4} & \textbf{84.0} \\
    Avg & 61.3 & 68.9 & \textbf{68.4} & \textbf{74.3} \\ \bottomrule
\end{tabular}
\label{tab:large}
\end{table}

Following Sentri~\cite{soro2022ask}, we employ a shared encoder~(\ie the parameters of the query encoder and the passage encoder are the same) with large size as the dual-encoder and evaluate our method. As shown in Table~\ref{tab:large}, \name with XLM-R Large achieves a significant performance improvement, which further demonstrates the effectiveness of the proposed \name.

\subsection{Effect of Scheduled Sampling}

Previously, we demonstrate the effectiveness of the scheduled sampling by ablation study. Here, we present two generated samples to qualitatively analyze the scheduled sampling. As shown in Table~\ref{tab:example}, for the first case, the generated queries in other languages have the same semantics as the query from the source language. The sample is effective in alignment training and is helpful to achieve better performance. For the second case, the generated query in Finnish is relevant to  the query from the source language but not synonymous. The sample is harmful to the model training. These samples indicate that the scheduled sampling is necessary for alignment training. In this way, we can reduce the impact of the cases which do not have the same semantics and further achieve better performance.

\begin{table}[t] \footnotesize
\centering
\setlength\tabcolsep{4pt}
\caption{Two generated examples. The span answers are in bold.}
\begin{tabular}{p{0.95\columnwidth}} \toprule

    \textbf{Passage:} Johanna Maria Magdalena "Magda" Goebbels (née Ritschel; \textcolor{red}{\textbf{11 November 1901 – 1 May 1945}}) was the wife of Nazi Germany\'s Propaganda Minister Joseph Goebbels. A prominent member of the Nazi Party, she was a close ally, companion and political supporter of Adolf Hitler. Some historians refer to her as the unofficial "First Lady" of Nazi Germany, while others give that title to Emmy Göring. \\ \midrule
    \textbf{Source Query~(Ja):}\begin{CJK}{UTF8}{min}ヨハンナ・マリア・マクダレナ・ゲッベルスは何歳で死去した？\end{CJK} \\
    \textbf{Translation:} At what age did Johanna Maria McDalena Goebbels die? \\  \midrule
    \textbf{Generated Query~(Ru):} \foreignlanguage{russian}{В каком возрасте умерла Магда Геббельс?} \\
    \textbf{Translation:} At what age did Magda Goebbels die? \\ \midrule
    \textbf{Generated Query~(Fi):} Minä vuonna Magda Goebbels kuoli? \\
    \textbf{Translation:} In what year did Magda Goebbels die? \\  \midrule

    \textbf{Passage:} Charles V \textcolor{red}{\textbf{(24 February 1500 – 21 September 1558)}} was ruler of both the Holy Roman Empire from 1519 and the Spanish Empire (as Charles V of Spain) from 1516, as well as of the lands of the former Duchy of Burgundy from 1506. He stepped down from these and other positions by a series of abdications between 1554 and 1556. Through inheritance, he brought together under his rule extensive territories in western, central, and southern Europe, and the Spanish viceroyalties in the Americas and Asia. \\  \midrule
    \textbf{Source Query~(Ko):} \begin{CJK}{UTF8}{mj}신성 로마 제국 카를 5세 재위기간은 얼마나 되나요?\end{CJK} \\
    \textbf{Translation:} How long was the reign of Charles V of the Holy Roman Empire? \\ \midrule
    \textbf{Generated Query~(Ru):} \foreignlanguage{russian}{Сколько лет правил Карл V?} \\
    \textbf{Translation:} How many years did Charles V rule? \\ \midrule
    \textbf{Generated Query~(Fi):} Minä vuonna Charles V hallitsi Rooman valtakuntaa? \\
    \textbf{Translation:} In what year did Charles V rule the Roman Empire? \\
    
    \bottomrule 
\end{tabular}
\label{tab:example}
\end{table}

\begin{table}[t] \footnotesize
\centering
\setlength\tabcolsep{4pt}
\caption{A generated example with different input templates. The span answer is in bold. Here, ``QG'' denotes query generator.}
\begin{tabular}{p{0.95\columnwidth}} \toprule

    \textbf{Passage:} The Higgs boson is an elementary particle in the Standard Model of particle physics, produced by the quantum excitation of the Higgs field, one of the fields in particle physics theory. It is named after physicist \textcolor{red}{\textbf{Peter Higgs}}, who in 1964, along with five other scientists, proposed the mechanism which suggested the existence of such a particle. Its existence was confirmed in 2012 by the ATLAS and CMS collaborations based on collisions in the LHC at CERN. \\ \midrule
    \textbf{Generated Query by \uline{QG w/ span answer}~(Fi):} Kuka on kehittänyt Higgs-boson? \\
    \textbf{Translation:} Who developed the Higgs boson? \\  \midrule
    \textbf{Generated Query by \uline{QG w/ span answer}~(Ko):} \begin{CJK}{UTF8}{mj}히그슨을 처음 발견한 사람은 누구인가?\end{CJK} \\
    \textbf{Translation:} Who first discovered Higson? \\ \midrule
    
    \textbf{Generated Query by \uline{QG w/o span answer}~(Fi):} Milloin Higgs on löydetty? \\
    \textbf{Translation:} When was Higgs Found? \\  \midrule
    \textbf{Generated Query by \uline{QG w/o span answer}~(Ko):} \foreignlanguage{russian}{Кто был первым исследователем физики Higgs?} \\
    \textbf{Translation:} Who was Higgs' first physics researcher? \\ \bottomrule 
\end{tabular}
\label{tab:answer-example}
\end{table}

\subsection{Effect of Span Answer}

In our method, we employ the span answer to encourage the query generator to generate synonymous queries. Here, we conduct experiments to evaluate the effect of the span answer. Specially, we use another template: \emph{``generate [language] query: [content]''} where we only need to fill two placeholders with the language of the target query and the passage content. We also incorporate the cross-encoder for comparison. We use the re-rankers to re-rank the retrieved results of the warm-up dual-encoder initialized with XLM-R. Note that introducing the span answer into the cross-encoder makes the re-ranking task easier, because the cross-encoder only needs to check whether the passage contains the span answer. The scores of this cross-encoder almost degenerate into hard labels and it is difficult to effectively train the dual-encoder by distilling knowledge from this cross-encoder.

We show the results in Table~\ref{tab:re-ranker}. Based on these results, we have the following findings.
On the one hand, the query generator trained with span answers is better than the query generator without span answers. It shows that taking span answers as input leads to better performance on re-ranking tasks for the query generator.
On the other hand, both the two query generator is better than the cross-encoder when re-ranking top-1000 retrieved passages, it shows the effectiveness of the query generator  in the cross-lingual setting. 

\begin{table*}[t] \footnotesize
\centering
\setlength\tabcolsep{4pt}
\caption{Performance comparison of different re-rankers on XOR-Retrieve dev set. The best results are in bold. Here, ``QG'' denotes query generator.}
\begin{tabular}{l|ccccccc|c|ccccccc|c} \toprule
    \multirow{2.5}{*}{Methods} & \multicolumn{8}{c|}{R@2kt} & \multicolumn{8}{c}{R@5kt} \\ \cmidrule(lr){2-17}
    ~ & Ar & Bn & Fi & Ja & Ko & Ru & Te & Avg & Ar & Bn & Fi & Ja & Ko & Ru & Te & Avg \\ \midrule
    Dual-Encoder & 35.3 & 43.1 & 50.3 & 35.7 & 44.6 & 31.2 & 50.0 & 41.5 & 49.5 & 54.9 & 59.2 & 45.2 & 55.1 & 31.2 & 63.4 & 53.4 \\ \midrule 
    \multicolumn{17}{c}{Re-ranking top-100 retrieved passages} \\ \midrule
    QG w/ Answer & \textbf{51.1} & \textbf{57.2} & 53.5 & \textbf{43.2} & \textbf{55.1} & \textbf{43.9} & 61.8 & \textbf{52.3} & \textbf{56.6} & 60.5 & 60.5 & \textbf{51.0} & \textbf{60.7} & 46.4 & 67.6 & \textbf{57.6} \\
    QG w/o Answer & 48.9 & 55.6 & 54.1 & 41.9 & 53.3 & 43.5 & 61.8 & 51.3 & \textbf{56.6} & \textbf{60.9} & 59.9 & 49.8 & 59.3 & 46.8 & \textbf{68.1} & 57.3 \\
    Cross-Encoder & 49.2 & 53.6 & \textbf{57.6} & 41.1 & 54.0 & 41.4 & \textbf{63.0} & 51.4 & 55.3 & 59.9 & \textbf{62.1} & 49.8 & 60.4 & \textbf{47.3} & 66.4 & 57.3 \\
     \midrule
    \multicolumn{17}{c}{Re-ranking top-1000 retrieved passages} \\ \midrule
    QG w/ Answer & \textbf{53.7} & \textbf{66.1} & \textbf{56.7} & \textbf{52.3} & \textbf{59.3} & \textbf{56.1} & \textbf{68.9} & \textbf{59.0} & \textbf{61.2} & \textbf{71.1} & 62.1 & \textbf{58.1} & 65.6 & \textbf{61.6} & \textbf{74.4} & \textbf{64.9} \\
    QG w/o Answer & 52.4 & 62.8 & 56.1 & 49.0 & 58.2 & 55.7 & 64.3 & 56.9 & 59.9 & 70.7 & 62.1 & 57.3 & 64.9 & 59.9 & 73.5 & 64.0 \\
    Cross-Encoder & 50.8 & 58.2 & 55.1 & 45.2 & \textbf{59.3} & 50.6 & 65.5 & 55.0 & \textbf{61.2} & 67.4 & \textbf{63.4} & 53.5 & \textbf{66.3} & 56.1 & 73.9 & 63.1 \\ 
    \bottomrule 
\end{tabular}
\label{tab:re-ranker}
\end{table*}

In addition, we show queries generated by the two query generators in Table~\ref{tab:answer-example}. As we can see, for the query generator that does not take the span answer as input, the generated queries can be answered by the passage, but they focus on different segments of the passage and they are not synonymous. On contrary, for the query generator that takes the span answer as input, generated queries can be answered by the passage and they are synonymous. It shows that taking the span answer as input can effectively encourage the generator to generate synonymous queries.

\begin{table*}[t] \footnotesize
\centering
\caption{Detailed performance on MKQA test set. ``$\ast$'' denotes that the results are copied from the Sentri paper.}
\begin{tabular}{ccccccccccc} \toprule
    Methods & Da & De & Es & Fr & He & Hu & It & Km & Ms & Nl \\ \midrule
    CORA$^{\ast}$ & 44.5 & 44.6 & 45.3 & 44.8 & 27.3 & 39.1 & 44.2 & 22.2 & 44.3 & 47.3 \\
    BM25 + MT$^{\ast}$ & 44.1 & 43.3 & 44.9 & 42.5 & 36.9 & 39.3 & 40.1 & 31.3 & 42.5 & 46.5 \\
    Sentri$^{\ast}$ & 57.6 & 56.5 & 55.9 & 55.1 & 47.9 & 51.8 & 54.3 & 43.9 & 56.0 & 56.3 \\
    ~~~ w/ Bi-Encoder$^{\ast}$ & 50.0 & 47.8 & 48.7 & 47.4 & 37.7 & 43.4 & 41.8 & 37.8 & 49.5 & 47.3 \\ \midrule
    \name & 58.3 & 56.4 & 55.2 & 55.5 & 44.7 & 52.4 & 52.3 & 42.0 & 56.9 & 57.5 \\
    \name w/ LaBSE & \textbf{63.3} & \textbf{61.8} & \textbf{62.2} & \textbf{62.4} & \textbf{56.1} & \textbf{58.9} & \textbf{60.6} & \textbf{53.0} & \textbf{64.2} & \textbf{63.0} \\ \midrule
    No & Pl & Pt & Sv & Th & Tr & Vi & Zh-cn & Zh-hk & Zh-tw & Avg \\ \midrule
    48.3 & 44.8 & 40.8 & 43.6 & 45.0 & 34.8 & 33.9 & 33.5 & 41.5 & 41.0 & 41.1 \\
    43.3 & 46.5 & 45.7 & 49.7 & 46.5 & 42.5 & 43.5 & 37.5 & 37.5 & 36.1 & 42.0 \\
    56.5 & 55.8 & 54.8 & 56.9 & 55.3 & 53.0 & 54.4 & 50.2 & 50.7 & 49.4 & 53.3 \\
    49.1 & 47.0 & 47.7 & 50.0 & 46.5 & 45.6 & 47.3 & 42.6 & 41.5 & 41.0 & 45.3 \\\midrule
    57.0 & 54.9 & 54.7 & 58.0 & 55.7 & 53.9 & 54.9 & 50.4 & 49.3 & 48.9 & 53.4 \\
    \textbf{62.8} & \textbf{62.0} & \textbf{61.5} & \textbf{63.3} & \textbf{60.5} & \textbf{60.6} & \textbf{61.8} & \textbf{57.3} & \textbf{56.3} & \textbf{56.0} & \textbf{60.3} \\ \bottomrule
\end{tabular}
\label{tab:mkqa-lang}
\end{table*}

\begin{table*}[t] \footnotesize
\centering
\setlength\tabcolsep{4pt}
\caption{Detailed performance for ablation study on XOR-Retrieve dev set. }
\begin{tabular}{l|ccccccc|c|ccccccc|c} \toprule
    \multirow{2.5}{*}{~} & \multicolumn{8}{c|}{R@2kt} & \multicolumn{8}{c}{R@5kt} \\ \cmidrule(lr){2-17}
    ~ & Ar & Bn & Fi & Ja & Ko & Ru & Te & Avg & Ar & Bn & Fi & Ja & Ko & Ru & Te & Avg \\ \midrule
    \name & 52.8 & \textbf{70.1} & 60.2 & \textbf{54.8} & \textbf{62.8} & \textbf{57.8} & \textbf{70.6} & \textbf{61.3} & \textbf{63.8} & \textbf{78.0} & \textbf{65.3} & \textbf{63.5} & \textbf{69.8} & \textbf{67.1} & 74.8 & \textbf{68.9} \\ \midrule
    w/o Sampling & \textbf{53.1} & 68.4 & 60.2 & 50.2 & 61.1 & 55.7 & 68.1 & 59.5 & 63.4 & \textbf{78.0} & 64.6 & 60.6 & 68.1 & 63.7 & 74.4 & 67.5 \\
    w/o Alignment & 51.8 & 68.1 & \textbf{60.8} & 51.0 & 60.4 & 57.4 & 70.2 & 59.9 & 61.5 & 74.7 & 64.6 & 62.7 & 68.8 & 62.4 & \textbf{75.2} & 67.1 \\
    w/o Generation & 55.0 & 68.1 & 59.6 & 47.3 & 60.7 & 54.9 & 66.4 & 58.8 & 62.8 & 74.0 & 65.0 & 56.4 & 66.3 & 62.0 & 74.8 & 65.9 \\
    w/o All & 35.3 & 43.1 & 50.3 & 35.7 & 44.6 & 31.2 & 50.0 & 41.5 & 49.5 & 54.9 & 59.2 & 45.2 & 55.1 & 31.2 & 63.4 & 53.4 \\ \bottomrule 
\end{tabular}
\label{tab:ablation-lang}
\end{table*}

\begin{table*}[t] \footnotesize
\centering
\setlength\tabcolsep{4pt}
\caption{Detailed performance of alignment based on different pre-trained languages models. }
\begin{tabular}{l|ccccccc|c|ccccccc|c} \toprule
    \multirow{2.5}{*}{~} & \multicolumn{8}{c|}{R@2kt} & \multicolumn{8}{c}{R@5kt} \\ \cmidrule(lr){2-17}
    ~ & Ar & Bn & Fi & Ja & Ko & Ru & Te & Avg & Ar & Bn & Fi & Ja & Ko & Ru & Te & Avg \\ \midrule
    XLM-R & \textbf{52.8} & \textbf{70.1} & 60.2 & \textbf{54.8} & \textbf{62.8} & \textbf{57.8} & \textbf{70.6} & \textbf{61.3} & \textbf{63.8} & \textbf{78.0} & \textbf{65.3} & \textbf{63.5} & \textbf{69.8} & \textbf{67.1} & 74.8 & \textbf{68.9} \\
    w/o Alignment & 51.8 & 68.1 & \textbf{60.8} & 51.0 & 60.4 & 57.4 & 70.2 & 59.9 & 61.5 & 74.7 & 64.6 & 62.7 & 68.8 & 62.4 & \textbf{75.2} & 67.1 \\ \midrule
    LaBSE & \textbf{67.3} & \textbf{78.9} & \textbf{65.9} & \textbf{59.8} & 66.3 & \textbf{63.7} & \textbf{80.7} & \textbf{68.9} & 72.2 & \textbf{83.2} & \textbf{69.7} & \textbf{68.0} & 70.9 & \textbf{71.7} & \textbf{84.9} & \textbf{74.4} \\
    w/o Alignment & 65.7 & 78.3 & 65.0 & 58.9 & \textbf{67.0} & 62.9 & 77.3 & 67.9 & \textbf{72.5} & 80.9 & \textbf{69.7} & 66.8 & \textbf{71.9} & 70.5 & 84.5 & 73.8 \\ \bottomrule 
\end{tabular}
\label{tab:align-lang}
\end{table*}

\section{Detailed Results}

Due to the limited space, we only present average performance for some experiments in Section~\ref{sec:exp}. Here, we present the detailed performance in all languages of these experiments. Firstly, we present the detailed performance of all methods on the MKQA test set in Table~\ref{tab:mkqa-lang}. Secondly, we present the detailed performance of ablation results in Table~\ref{tab:ablation-lang}. Finally, we present the detailed performance for evaluating the effect of alignment in Table~\ref{tab:align-lang}.

\end{document}